\documentclass[prd,twocolumn,preprintnumbers]{revtex4-1}

\usepackage{amsmath}
\usepackage{amsfonts}
\usepackage{epsfig}
\usepackage{graphicx}
\usepackage{color}
\usepackage[normalem]{ulem}
\usepackage{url}
\usepackage{float}
\usepackage[breaklinks, plainpages=false, colorlinks=true, anchorcolor=cyan, linkcolor=red, citecolor=cyan, urlcolor=magenta, bookmarks=false]{hyperref}

\usepackage[caption=false]{subfig}

\usepackage{aas_macros}

\begin{document}
\renewcommand{\thefigure}{\arabic{figure}}
\setcounter{figure}{0}

 \def\I{{\rm i}}
 \def\E{{\rm e}}
 \def\D{{\rm d}}

\bibliographystyle{apsrev}

\title{Escaping Neal's Funnel: a multi-stage sampling method for hierarchical models}

\author{Aiden Gundersen}
\affiliation{eXtreme Gravity Institute, Department of Physics, Montana State University, Bozeman, Montana 59717, USA}

\author{Neil J. Cornish}
\affiliation{eXtreme Gravity Institute, Department of Physics, Montana State University, Bozeman, Montana 59717, USA}

\begin{abstract} 
Neal's funnel refers to an exponential tapering in probability densities common to Bayesian hierarchical models. Usual sampling methods, such as Markov Chain Monte Carlo, struggle to efficiently sample the funnel. Reparameterizing the model or analytically marginalizing local parameters are common techniques to remedy sampling pathologies in distributions exhibiting Neal's funnel. In this paper, we show that the challenges of Neal's funnel can be avoided by performing the hierarchical analysis, well, hierarchically. That is, instead of sampling all parameters of the hierarchical model jointly, we break the sampling into multiple stages. The first stage samples a generalized (higher-dimensional) hierarchical model which is parameterized to lessen the
sharpness of the funnel. The next stage samples from the estimated density of the first stage, but under a constraint which restricts the sampling to recover the marginal distributions on the hyper-parameters of the original (lower-dimensional) hierarchical model. A normalizing flow can be used to represent the distribution from the first stage, such that it can easily be sampled from for the second stage of the analysis. This technique is useful when effective reparameterizations are computationally expensive to calculate, or a generalized hierarchical model already exists from which it is easy to sample.
\end{abstract}

\maketitle

\section{Introduction}\label{sec:intro}
Neal’s funnel is an exponential tapering of volumes in parameter space containing significant probability mass. These geometries are a common feature of Bayesian hierarchical models, and it's difficult to sample from such densities~\cite{NealsFunnel}. The funnel is caused by a (global) hyper-parameter controlling the variance of a (lower-level) local parameter in the prior. When the hyper-parameter suppresses the variance of a local parameter the throat of the funnel is formed. The throat is generally small, but dense, containing significant probability mass. The opening of the funnel is formed where the local parameter is allowed to have a large variance by the hyper-parameter. Standard Markov Chain Monte Carlo (MCMC) methods often struggle to sample the throat of the funnel, as precise jump proposals must be made along the funnel, and adapted to the rapidly changing covariance. When sampling such hierarchical models, naive samplers achieve low sampling rates or fail to explore the full parameter space, often getting ``stuck" in the throat of the funnel.

There are various ways to improve convergence times when sampling densities exhibiting Neal’s funnel. A common practice is to consider the local parameters of the hierarchical model nuisance parameters, and analytically marginalize them out of the model, if possible. However, analytic marginalization can render the target density computationally expensive to evaluate, bottle-necking the sampling if the marginalized density remains high-dimensional. Nonetheless Neal’s funnel is marginalized out of the model with this approach, and a target density of simpler geometry remains. Another approach is to transform the parameters of the model so that they can be sampled from simpler distributions~\cite{reparamFunnel}. Samplers which fail to converge on distributions with Neal’s funnel may become well-behaved after a clever reparameterization, or obtain effective samples at a higher rate. Lastly a more robust sampler can handle Neal’s funnel: Riemannian Manifold Hamiltonian Monte Carlo, which uses a position-dependent metric, has proven effective at sampling Neal’s funnel in hierarchical models~\cite{RMHMC1, RMHMC2}.

In this paper, we present another potential avenue for working around Neal's funnel.
Consider an arbitrary hierarchical model, observed data $d$, local parameters $\mathbf{x}$, and hyper-parameters $\mathbf{y}$, where $\text{dim}(\mathbf{x})=n$ and $\text{dim}(\mathbf{y})=m$. The joint posterior density is
\begin{equation}\label{eq:HierarchINTRO}
    p(\mathbf{x}, \mathbf{y}|d)\propto p(d | \mathbf{x})\cdot p(\mathbf{x}|\mathbf{y})\cdot p(\mathbf{y})\,,
\end{equation}
where $p(d|\mathbf{x})$ is the likelihood which depends only on the local parameters and data observed, $p(\mathbf{x}|\mathbf{y})$ is the prior on the local parameters, conditioned by the hyper-parameters, and $p(\mathbf{y})$ is the hyper-prior. If the hyper-parameters $\mathbf{y}$ parameterize the variance of $\mathbf{x}$ in the conditional prior, $p(\mathbf{x}|\mathbf{y})$, the posterior density may be difficult to sample because of Neal's funnel. Our goal is to obtain samples from the marginalized distribution on the hyper-parameters $p(\mathbf{y}|d)$, while mitigating the sampling difficulties of the funnel. Rather than analytically marginalizing over the local parameters $\mathbf{x}$ or implementing reparameterizations, our approach breaks the sampling into multiple stages. We first sample from a hierarchical model which uses a higher-dimensional generalized hyper-model,
\begin{equation}\label{eq:genHierarchINTRO}
    p(\mathbf{x}, \mathbf{z}|d)\propto p(d | \mathbf{x})\cdot p(\mathbf{x}|\mathbf{z})\cdot p(\mathbf{z})\,,
\end{equation}
where $\mathbf{z}$ are  hyper-parameters of a generalized, higher-dimensional hyper-model, $\text{dim}(\mathbf{z})= M > m$. That is, there exists a mapping $\mathbf{z}=\mathbf{z}(\mathbf{y})$ which is injective, but non-surjective, and takes parameters of the original hyper-model into the space of the generalized hyper-model. This generalized hierarchical model, if chosen carefully, may reduce the sharpness of the funnel, and is therefore easier to sample with standard techniques.

Once Eq.~(\ref{eq:genHierarchINTRO}) is sufficiently sampled, the local parameters may be numerically marginalized over to obtain the distribution on the generalized hyper-parameters,
\begin{equation}\label{eq:marginalization}
    p(\mathbf{z}|d) = \int p(\mathbf{x}, \mathbf{z} | d)\,d\mathbf{x} \approx \frac{1}{N}\sum_{i=1}^N
\delta(\mathbf{z} - \mathbf{z}_i)
\end{equation}
where $\delta$ is the Dirac delta function and $\mathbf{z}_i$ is the $i^\text{th}$ sample of $N$ total samples. In order to obtain the marginal distribution on the original hyper-parameters, we estimate the density of the generalized (numerically) marginalized distribution,
\begin{equation}\label{eq:NF}
    \hat{p}_d(\mathbf{z}) \approx p(\mathbf{z}|d)\,.
\end{equation}
such that $\hat{p}_d(\mathbf{z})$ is the probability density function estimated from the marginalized samples of $p(\mathbf{z}|d)$. The marginalized distribution on the original hyper-parameters, $p(\mathbf{y}|d)$, is determined by the mapping from the original to the generalized hyper-model,
\begin{equation}\label{eq:constraint}
    p(\mathbf{y}|d)\approx \hat{p}_d(\mathbf{z}(\mathbf{y}))\cdot p(\mathbf{y})\,.
\end{equation}
That is, the estimated density, $\hat{p}_d$, can be sampled under the constraint which restricts the generalized hyper-model to the sub-surface of the original hyper-model, while remembering to include the original hyper-prior, $p(\mathbf{y})$.

In summary, to avoid Neal's funnel present in Eq.~(\ref{eq:HierarchINTRO}), we break the sampling into three steps:
\begin{enumerate}
    \item Sample a hierarchical model with a generalized (higher-dimensional) hyper-model, Eq.~(\ref{eq:genHierarchINTRO}).
    \item Estimate the density of the generalized hyper-parameter samples, Eq.~(\ref{eq:NF}).
    \item Sample the estimated density under a constraint which recovers the marginal distribution of the original hyper-parameters, Eq.~(\ref{eq:constraint}).
\end{enumerate}
In the generalized hyper-model, the dimension is increased and the hyper-parameters transformed to lessen the sharpness of the funnel. Samplers ill-equipped to handle Neal's funnel will learn this generalized density more efficiently if the generalization smooths out the funnel. The local parameters are sampled jointly with the generalized hyper-parameters at this stage, enabling us to numerical marginalize the local parameters from the model. The density is then estimated for the generalized hyper-parameter samples. Lastly, we sample the estimated density under the constraint of the original hyper-model. This last sampling step importantly includes the weighting from hyper-priors and determinants of Jacobians of transformations of previous steps.

Geometrically, the idea is to generalize the original hyper-model to a higher-dimensional hyper-volume, in which the original model is some embedded lower-dimensional surface. Numerically marginalizing the local parameters is easier in this hyper-volume, where the troubles of Neal's funnel are mitigated. Once the density of the generalized hyper-parameters is estimated, the model is resampled under the constraint that we only explore the embedded surface of the original hyper-model, recovering the marginal distribution of the original hyper-parameters.

Any density estimation technique is amenable to the methods presented in this paper. Here we will estimate the density of samples using normalizing flows (for a review, see e.g.~\cite{flowsReview}). Moreover, this multi-stage sampling approach may be extended to models with arbitrary levels of hierarchical coupling. The same three steps are repeated for every level and its hyper-coupling. In this paper, our examples are restricted to two level hierarchies for simplicity.

We compare the multi-stage sampling (MSS) method presented in this paper with a few other sampling schemes: naive sampling (NS), prior reparameterized sampling (PRS), and conditional posterior reparameterized sampling (CPRS). NS feeds the target density directly into the sampler under no reparameterization. PRS samples the parameters first from a standard normal distribution, then colors them using a Cholesky decomposition of the covariance of the prior. CPRS samples the local parameters first from a standard normal distribution, then colors them with the Cholesky decomposition of the covariance of the posterior, conditioned on the hyper-parameters. CPRS is a powerful sampling technique which was derived and implemented in code for the authors by R. van Haasteren~\cite{decentering}.

All probability densities are implemented in \texttt{JAX}~\cite{JAX} and sampled using a Hamiltonian Monte Carlo (HMC) scheme~\cite{HMC1, HMC2, HMC3} with the \texttt{NumPyro} package and its No U-Turn Sampler (NUTS)~\cite{NumPyro1, NumPyro2}. We estimate densities using normalizing flows implemented in the \texttt{FlowJAX} software~\cite{flowJAX}. In Section~\ref{sec:Neal's Funnel} we apply these sampling methods to the ``classic" Neal's funnel presented in Ref.~\cite{NealsFunnel}. MSS is trivial for this first example because it lacks a likelihood factor. Section~\ref{sec:Extended Funnel} demonstrates MSS on a less trivial funnel. Section~\ref{sec:RN example} introduces and analyzes a problem from gravitational wave astronomy, which inspired the MSS method presented here. Future applications are discussed in Section~\ref{sec:Future work}.

\section{Classic Neal's Funnel}\label{sec:Neal's Funnel}
We demonstrate our multi-stage sampling (MSS) approach on the classic funnel presented in Ref.~\cite{NealsFunnel}. The hierarchical model supports local parameters $\mathbf{x}\in\mathbb{R}^9$ and a hyper-parameter $y\in\mathbb{R}$ such that the joint distribution is
\begin{align}\label{eq:NealHierarch}
    p(\mathbf{x}, y) &= p(y) \cdot p(\mathbf{x}|y) \nonumber \\
    &= p(y) \cdot \prod_{i=1}^9 p(x_i|y)   
\end{align}
where
\begin{eqnarray}\label{eq:NealDists}
    y\sim&\mathcal{N}(\mu=0, \sigma=3) \nonumber \\
    x_i|y\sim&\mathcal{N}(\mu=0, \sigma=e^{y/2})\,.
\end{eqnarray}
The index $i \in\{1,2,\dots,9\}$ labels the components of the local parameter $\mathbf{x}$. Geometrically, this hierarchical model is a 10-dimensional funnel. If $y$ is negative, the conditional distribution on $\mathbf{x}$ narrows and the throat of the funnel is formed. Naive samplers will struggle to explore the distribution efficiently for negative values of $y$.

It is easy to reparameterize this model so the sampling is trivial. Rather than sampling $\mathbf{x}$ and $y$ directly, one should sample random variables $\hat{\mathbf{x}}$ and $\hat{y}$ from the standard normal distribution. These standardized random variables can be transformed to the parameters of interest with a simple rescaling: $y = 3\hat{y}$ and $\mathbf{x} = e^{y/2}\,\hat{\mathbf{x}}$. Alternatively, we can use a multi-stage sampling scheme described below.

The first step of our method is to generalize the hierarchical model to a higher-dimensional parameter space from which it is easier to sample. For this example, our generalized hierarchical model is
\begin{align}\label{eq:GenHierarch}
    p(\mathbf{x}, \mathbf{z}) &= p(\mathbf{x}|\mathbf{z})\cdot p(\mathbf{z}) \nonumber \\
    &= \prod_{i=1}^9 p(x_i|z_i) \cdot p(z_i)
\end{align}
where 
\begin{align}\label{eq:GenDists}
    \log_{10}z_i\sim\text{Uniform}(-a, a) \nonumber \\
    x_i|z_i\sim\mathcal{N}(\mu=0,\sigma=z_i)
\end{align}
where the bound $a = 4$ is chosen to sufficiently capture the probability mass of the hyper-prior in the original model, Eq.~(\ref{eq:NealDists}). This generalized hierarchical model is nearly twice the dimension of the previous model. However, when sampling over $\log_{10}\mathbf{z}$ the parameterization of the covariance matrix of the conditional distribution is such that the sharpness of the funnel is lessened and the generalized hierarchical model is easier to sample with usual methods. Moreover, this generalized model may be constrained to recover the original model under the mapping
\begin{equation}\label{eq:constraintNEALS}
    z_i = z_i(y) = e^{y/2}\hspace{3mm}\Leftrightarrow\hspace{3mm}\log_{10}z_i = \frac{y}{2}\,\log_{10}e
\end{equation}
for every $i\in\{1,2,\dots,9\}$.

Once the generalized distribution, Eq.~(\ref{eq:GenHierarch}), is sampled with usual MCMC methods, the next step is to learn the distribution of the generalized hyper-parameter samples numerically marginalized over the local parameters, $p(\log_{10}\mathbf{z})$. We learn the distribution with a normalizing flow which provides a density estimation for the samples,
\begin{equation}\label{eq:NFNEALS}
    \hat{p}(\log_{10}\mathbf{z})\approx p(\log_{10}\mathbf{z}) = \int p(\mathbf{x},\log_{10}\mathbf{z})\,d\mathbf{x}\,.
\end{equation}
Then we may sample this estimated density, $\hat{p}(\mathbf{z})$, in the original hyper-parameter, $y$, under the mapping of Eq.~(\ref{eq:constraintNEALS}), while remembering to impose the hyper-prior of the original model,
\begin{equation}\label{eq:stage2NEALS}
    p(y) \propto \hat{p}(\log_{10}\mathbf{z}(y))\cdot p(y)
\end{equation}
Assuming the density $\hat{p}(\log_{10}\mathbf{z})$ was accurately estimated with the normalizing flow, this resampling will recover the marginal distribution on the hyper-parameter $y$.

\begin{figure}
    \centering
    \includegraphics[width=0.9\linewidth]{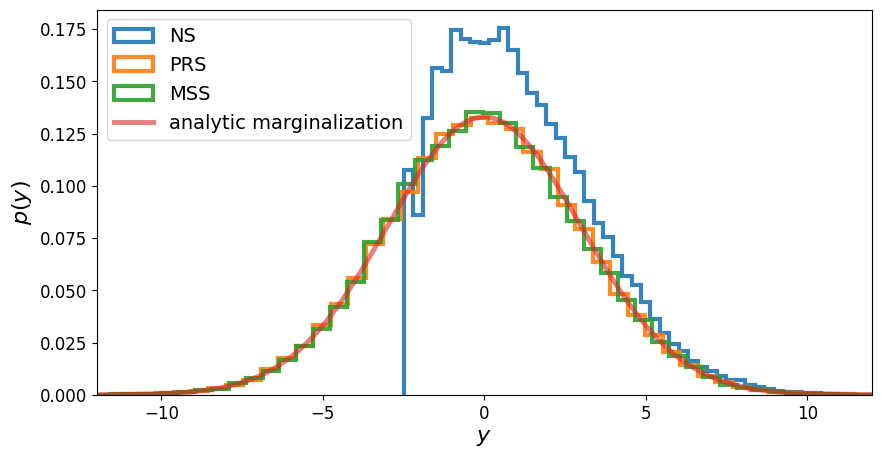}
    \caption{Marginalized distributions of the hyper-parameter $y$, under the hierarchical model of Eq.~(\ref{eq:NealHierarch}). The blue, orange, and green histograms use naive sampling, prior reparameterized sampling, and multi-stage sampling respectively. The red curve is the hyper-prior on $y$ from Eq.~(\ref{eq:NealDists}) and equivalently the analytically marginalized distribution on $y$, for this example.}
    \label{fig:Neals_marginal}
\end{figure}

In Figure~\ref{fig:Neals_marginal} we show the marginal distribution on the hyper-parameter $y$ recovered using NS, PRS, and MSS. The naive sampler gets stuck in the throat of the funnel and fails to explore the entire parameter space. Multi-stage sampling however escapes the difficulties of the funnel and is consistent with the analytically marginalized distribution, and the distribution obtained using standard reparameterization techniques.

Multi-stage sampling is able to recover the marginal distribution on the hyper-parameter better than naive sampling, which gets stuck in the throat of the funnel. However, this example is trivial because there is no \textit{tension} between the prior and a likelihood in the model. In other words, our hierarchical model is completely prior dominated and the marginal distribution on the hyper-parameter is precisely the hyper-prior itself. The means the estimated density Eq.~(\ref{eq:NFNEALS}) is uniform and Eq.~(\ref{eq:stage2NEALS}) is trivial. In the next section we add a likelihood factor which distinguishes the hyper-prior from the marginal distribution on the hyper-parameter.

\section{Neal's Funnel with a Likelihood}\label{sec:Extended Funnel}

We repeat the analysis of the last section, now including a likelihood term in the hierarchical model. Assume again the model supports local parameters $\mathbf{x}\in\mathbb{R}^9$ and a hyper-parameter $y\in\mathbb{R}$. Additionally, we have observed some data $d$ modeled by the local parameters so the posterior density of the hierarchical model is
\begin{align}\label{eq:LikeFunnelHierarch}
    p(\mathbf{x}, y | d) &\propto p(d | \mathbf{x})\cdot p(\mathbf{x}|y) \cdot p(y) \nonumber \\
    &\propto p(y) \prod_{i=1}^9 p(d|x_i)\cdot p(x_i|y)
\end{align}
where
\begin{eqnarray}\label{eq:LikeFunnelDists}
    d|x_i\sim&\mathcal{N}(\mu=2, \sigma=5) \nonumber \\
    y\sim&\mathcal{N}(\mu=0, \sigma=3) \nonumber \\
    x_i|y\sim&\mathcal{N}(\mu=0, \sigma=e^{y/2})\,,
\end{eqnarray}
and $i\in\{1,2,...,9\}$ indexes the components of the local parameter $\mathbf{x}$.

The same generalized hierarchical model is used as in the previous section, Eq.~(\ref{eq:GenHierarch}) and Eq.~(\ref{eq:GenDists}), and the same equation of constraint relates the generalized and original hyper-model, Eq.~(\ref{eq:constraintNEALS}). The only difference here is we include the likelihood weighting while sampling the generalized model. This means the marginalized distribution on the generalized hyper-parameters, which is learned by the normalizing flow, is now conditioned on the data,
\begin{equation}\label{eq:NFNEALSLIKE}
    \hat{p}_d(\log_{10}\mathbf{z})\approx p(\log_{10}\mathbf{z}|d) = \int p(\mathbf{x},\log_{10}\mathbf{z}|d)\,d\mathbf{x}\,.
\end{equation}
Importantly, the conditional on the data distinguishes this non-trivial estimated density from the hyper-prior of Eq.~(\ref{eq:LikeFunnelDists}), as opposed to the example of Section~\ref{sec:Neal's Funnel}. The estimated density is resampled in $y$ under the constraint of Eq.~(\ref{eq:constraintNEALS}) so that marginalized distribution on the original hyper-parameter is
\begin{equation}\label{eq:stage2NEALSLIKE}
    p(y|d) \propto \hat{p}(\log_{10}\mathbf{z}(y)|d)\cdot p(y)\,.
\end{equation}

\begin{figure}
    \centering
    \includegraphics[width=0.9\linewidth]{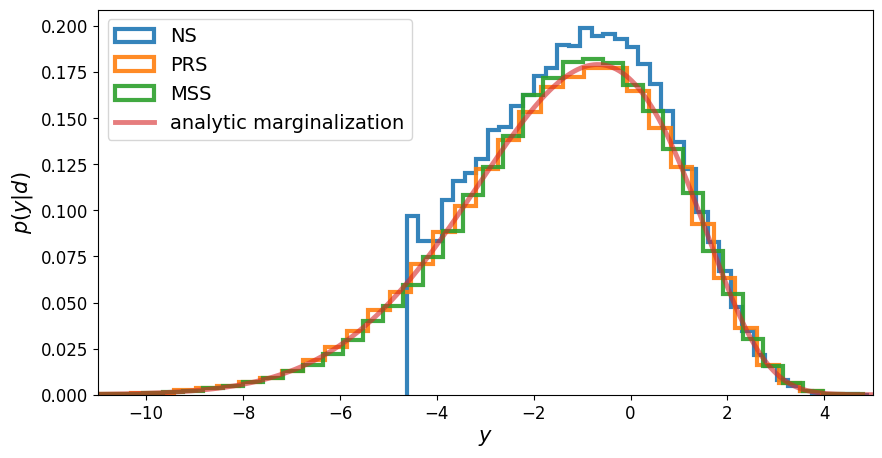}
    \caption{Marginalized distributions of the hyper-parameter $y$, under the hierarchical model of Eq.~(\ref{eq:LikeFunnelHierarch}). The blue, orange, and green histograms use naive sampling, prior reparameterized sampling, and multi-stage sampling respectively. The red curve is the posterior on $y$, analytically marginalized over the local parameters.}
    \label{fig:like_marginal}
\end{figure}

The recovered marginal distributions on the hyper-parameter $y$ are shown in Figure~\ref{fig:like_marginal}. Naive MCMC sampling again gets stuck in the throat of the funnel. Standardizing the sampled parameters under a reparameterization of the (hyper-) prior distribution is again a viable option to explore the parameter space. Multi-stage sampling is consistent with the distribution obtained using reparameterization, illustrating the viability of MSS.

This example is importantly distinguished from the last section by the presence of the likelihood. This renders the hyper-prior and marginal distributions on the hyper-parameter distinct, $p(y|d)\neq p(y)$. Then the normalizing flow estimates a non-trivial distribution, Eq.~(\ref{eq:NFNEALSLIKE}), which is to be resampled under a constraint, Eq.~(\ref{eq:constraintNEALS}), such that the original marginal distribution is recovered. When the hyper-prior and marginal distribution on the hyper-parameter are identical, as in the previous example, the distribution resampled using the normalizing flow is precisely the hyper-prior of the generalized model, which is redundant because we already have a functional implementation of that density.

\section{Time-domain analyses of stochastic red processes}\label{sec:RN example}

In this section we present a simplified version of a problem from gravitational wave astronomy, which inspired the multi-stage sampling method presented in this paper. Hierarchical models in gravitational wave data analysis can exhibit Neal's funnel which challenges the sampling, particularly in pulsar timing arrays. Pulsar timing arrays (PTAs) measure the time-of-arrival (TOAs) of radio pulses produced by millisecond pulsars. By recording these TOAs for decades, PTAs are sensitive to gravitational waves (GWs) of nano-Hertz frequencies. Presently, evidence for a stochastic gravitational wave background (GWB) has been found by the North American Nanohertz Observatory for Gravitational Waves, the European Pulsar Timing Array, the Parkes Pulsar Timing Array, and the Chinese Pulsar Timing Array~\cite{NG15, EPTAGWB, ParkesPTA, ChinesePTA}. For an introduction to PTA science, see e.g.~\cite{nHzGWastro}.

One goal of PTA data analysis is to understand the spectral properties of the dataset, as physical models often make predictions for the spectra of say, a gravitational wave background~\cite{13/3}. However, we are unable to perform a Fourier transform in this context because the data is unevenly sampled in the time-domain, and the detector noise is heteroscedastic. Instead we perform the Bayesian analogue of a Fourier transform by modeling a set of Fourier coefficients which describe the time-domain data. Spectral properties of the data can then be inferred via a hierarchical Bayesian model.

The full PTA data analysis pipeline consists of many complex signal and noise models. Here we present a simplified single pulsar analysis, which models a red stochastic process. This simplified model serves to illustrate Neal's funnel as it appears in PTA data analysis, and the multi-stage sampling approach which can circumvent the difficulties of a funnel.

Consider $N$ discrete time-domain data $\mathbf{d} = (d_1, d_2, ..., d_N)^\text{T}$ sampled over (unevenly spaced) ordered time samples $\mathbf{t} = (t_1, t_2, ..., t_N)^\text{T}$. A simplified PTA dataset consists of a stochastic red signal, $\mathbf{s}_R$, and white noise, $\mathbf{n}_W$, such that
\begin{equation}
    \mathbf{d} = \mathbf{s}_R + \mathbf{n}_W\,.
\end{equation}
The white noise is independent and identically distributed over the time samples according to a zero mean one-dimensional normal distribution, whose variance known. That is, the white noise covariance matrix is $\mathbf{N} = \langle \mathbf{n}\mathbf{n}^\text{T}\rangle = \sigma^2\,\mathbf{I}$. The red process obeys some spectral model, with power predominately in the low frequency bins, and is parameterized by a set of Fourier coefficients. These coefficients are mapped to the time-domain via $\mathbf{s}_R = \mathbf{F}\mathbf{a}$ where $\mathbf{a} = (a_1, b_1, a_2, b_2, ..., a_{N_f}, b_{N_f})^\text{T}$ are the Fourier coefficients which parameterize the red signal, $N_f$ is the number of frequency bins modeled, $\mathbf{F}$ is the Fourier design matrix defined as
\begin{widetext}
\begin{equation}
\mathbf{F} = \begin{pmatrix}
    \sin(2\pi t_1/T) & \cos(2\pi t_1/T) & ... & \sin(2\pi N_f t_1/T) & \cos(2\pi N_f t_1/T) \\
    \sin(2\pi t_2/T) & \cos(2\pi t_2/T) & ... & \sin(2\pi N_f t_2/T) & \cos(2\pi N_f t_2/T) \\
    \vdots & \vdots & \vdots & \vdots & \vdots \\
    \sin(2\pi t_N/T) & \cos(2\pi t_N/T) & ... & \sin(2\pi N_f t_N/T) & \cos(2\pi N_f t_N/T) \\
\end{pmatrix}\,,
\end{equation}
\end{widetext}
and $T=t_N - t_1$ is the observational period of the data.

Subtracting the stochastic red signal from the data we are left with a particular realization of white noise, so the likelihood, up to a multiplicative constant, is
\begin{equation}\label{eq:PTA_like}
    p(\mathbf{d}|\mathbf{a}) \propto \text{exp}\bigg[-\frac{1}{2}(\mathbf{d} - \mathbf{F}\mathbf{a})^\text{T}\,\mathbf{N}^{-1}\,(\mathbf{d} - \mathbf{F}\mathbf{a})\bigg]\,.
\end{equation}
Hierarchically, we model the Fourier coefficients of the red signal as a Gaussian process such that $\mathbf{a}\sim\mathcal{N}(\mathbf{0}, \boldsymbol{\phi})$ and the covariance is parameterized according to some spectral model with hyper-parameters $\boldsymbol{\eta}$, i.e., $\boldsymbol{\phi} = \boldsymbol{\phi}(\boldsymbol{\eta})$. 
The prior on Fourier coefficients is
\begin{equation}\label{eq:PTA_prior}
    p(\mathbf{a}|\boldsymbol{\eta}) = \frac{1}{\sqrt{\text{det}(2\pi\boldsymbol{\phi})}}\,\text{exp}\bigg[-\frac{1}{2}\mathbf{a}^\text{T}\,\boldsymbol{\phi}^{-1}\,\mathbf{a}\bigg]\,.
\end{equation}
Up to a normalization constant, the hierarchical posterior density is
\begin{equation}\label{eq:PTApost}
    p(\mathbf{a}, \boldsymbol{\eta}|\mathbf{d}) \propto p(\mathbf{d}|\mathbf{a})\cdot p(\mathbf{a}|\boldsymbol{\eta})\cdot p(\boldsymbol{\eta})
\end{equation}
where $p(\boldsymbol{\eta})$ is the hyper-prior.

A common spectral model is a power law which models the covariance of the Fourier coefficients as a diagonal matrix with elements
\begin{equation}\label{eq:power law}
    \text{diag}(\boldsymbol{\phi}) \,(A, \gamma) = A\,\bigg(\frac{f_i}{f_\text{ref}}\bigg)^{-\gamma}
\end{equation}
where $i\in\{1, 2, \dots, N_f\}$, the hyper-parameters $\boldsymbol{\eta}=(A, \gamma)$ are the amplitude and spectral index of the power law, respectively, and $f_\text{ref}$ is a reference frequency. Conventionally, a log-uniform hyper-prior is placed on the amplitude of the power law and a uniform hyper-prior is placed on the spectral index, $\gamma$.

In standard PTA analyses, the posterior density Eq.~(\ref{eq:PTApost}) is simplified by expressing Fourier coefficients as random variables of a multivariate normal distribution, then analytically marginalizing them out of the model~\cite{Lentati}. This reduces the dimension of the parameter space, but the marginalized posterior contains a dense covariance matrix which must be inverted at every evaluation, rendering the posterior evaluation computationally expensive. Recent work has opted to directly sample the coefficients instead of analytically marginalizing them out of the model, retaining the posterior in the form of Eq.~(\ref{eq:PTApost}),~\cite{Lentati, Laal2023, Laal2025, vanHaasteren, gundersen2024rapid}. This increases the dimension of the parameter space, but the posterior evaluation is hyper-efficient.

\begin{figure*}
    \centering
    \includegraphics[width=0.8\linewidth]{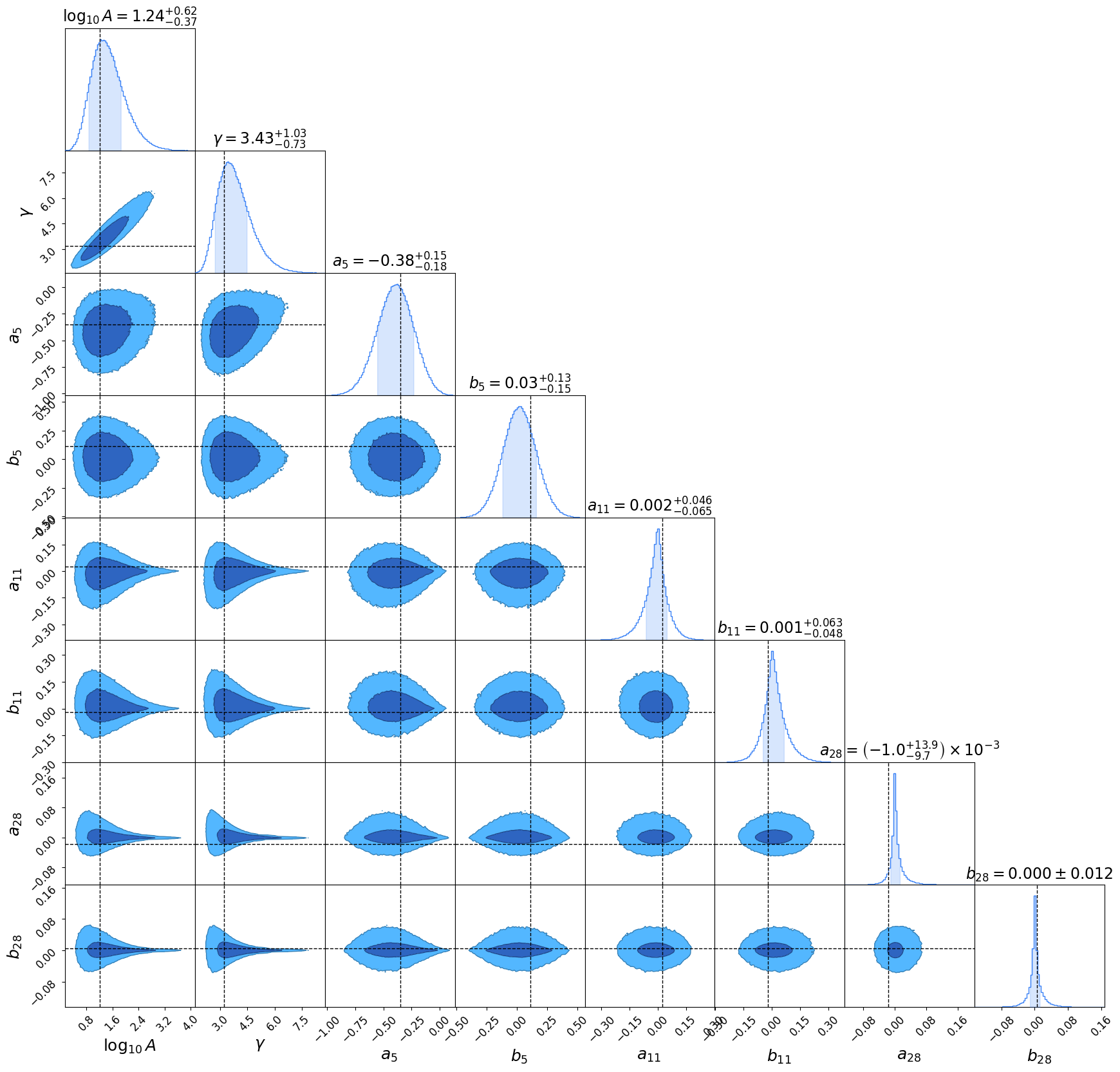}
    \caption{Samples from the PTA hierarchical model in Eq.~(\ref{eq:PTApost}). The first two columns are the power law hyper-parameters, amplitude and spectral index respectively. The remaining columns are the Fourier coefficients whose subscript denotes the frequency bin. The ``a" coefficients scale sine functions and ``b" cosine. The summary statistics and shading of the one-dimensional marginal distributions denote the maximum posterior and $1\sigma$ either side. The 1 and $2\sigma$ contours are shown in the two-dimensional distributions. Neal's funnel is observed between the hyper-parameters and Fourier coefficients in higher frequency bins.}
    \label{fig:funnel}
\end{figure*}

Despite hyper-efficient posterior evaluations, Eq.~(\ref{eq:PTApost}) is still difficult to sample directly because of Neal's funnel as illustrated in Figure~\ref{fig:funnel}. The power law spectral model (applied as a hyper-model) imposes a funneling geometry on the coefficients. When the amplitude of the power law is large (or the spectral index small), each frequency bin is allowed significant power. The variance of the coefficients is large in this case. When the amplitude is small (or the spectral index large), the power in each frequency bin is small and conditional distributions on the Fourier coefficients have small variance, forming the throat of the funnel. Naive samplers in PTA analyses need to analytically marginalize over the coefficients, or they will get stuck and fail to explore the parameter space effectively.

In order to conduct multi-stage sampling for this example, we generalize the power law hyper-model to a \textit{free spectral model}. A free spectral model allows the power to vary freely in each frequency bin. That is, each diagonal element of the prior covariance matrix is itself a free parameter: $\text{diag}(\boldsymbol{\phi})=\boldsymbol{\rho}\in\mathbb{R}_+^{N_f}$. The generalized hierarchical model is
\begin{equation}\label{eq:PTApostFS}
    p(\mathbf{a}, \boldsymbol{\rho}|\mathbf{d}) \propto p(\mathbf{d}|\mathbf{a})\cdot p(\mathbf{a}|\boldsymbol{\rho})\cdot p(\boldsymbol{\rho})\,.
\end{equation}
The sharpness of Neal's funnel is reduced when sampling over $\log_{10}\boldsymbol{\rho}$ in the higher-dimensional free spectral model. The free spectral model is useful in its own right, being a tool to assess alternative spectral models or exotic signals in PTA datasets~\cite{NG15_new_physics}. Here, it serves as the first stage of our multi-stage sampling method.

After sampling the generalized hierarchical model, Eq.~(\ref{eq:PTApostFS}), the marginalized distribution of the (generalized) free spectral hyper-parameters is learned with a normalizing flow,
\begin{equation}
    \hat{p}_\mathbf{d}(\log_{10}\boldsymbol{\rho}) \approx p(\log_{10}\boldsymbol{\rho}|d) = \int p(\log_{10}\boldsymbol{\rho}, \mathbf{a}|\mathbf{d})\,d\mathbf{a}\,.
\end{equation}
The constraint equation which restricts the (generalized) free spectral to the power law hyper-model is
\begin{equation}\label{eq:constraintPTA}
    \rho_i(A, \gamma) = A\bigg(\frac{f_i}{f_\text{ref}}\bigg)^{-\gamma}
\end{equation}
where $i\in\{1,2,..., N_f\}$. The marginalized distribution on the power law hyper-parameters can be recovered under the mapping,
\begin{equation}
    p(A, \gamma | \mathbf{d}) \approx \hat{p}_\mathbf{d}(\log_{10}\boldsymbol{\rho}(A, \gamma))\cdot p(A, \gamma)\,.
\end{equation}

\begin{figure}
    \centering
    \includegraphics[width=0.8\linewidth]{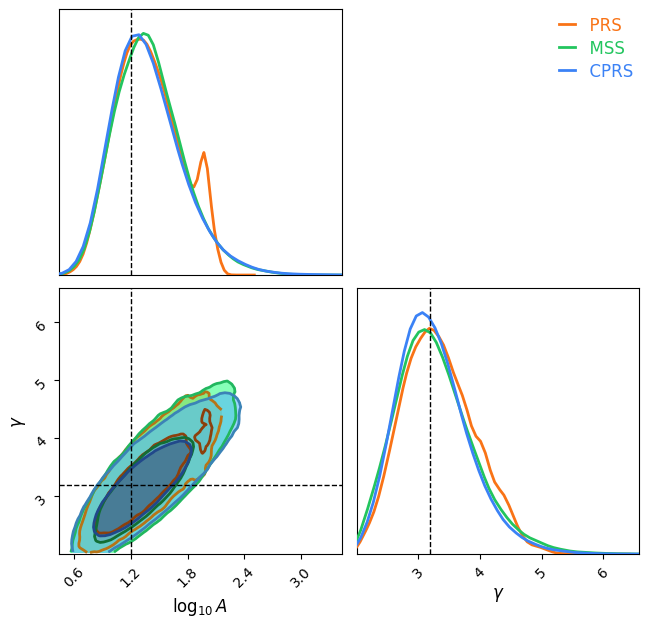}
    \caption{The distribution of power law hyper-parameters from analysis of a simplified PTA dataset. The orange, green, and blue distributions show prior reparameterized sampling, multi-stage sampling, and conditional posterior reparameterized sampling respectively. The contours of the two-dimensional distribution enclose the 1 and $2\sigma$ regions.}
    \label{fig:power_law}
\end{figure}

The distribution on the power law hyper-parameters obtained through various sampling methods is shown in Figure~\ref{fig:power_law}. The PRS method struggles with Neal's funnel in this example, and we must go further with CPRS to learn the density adequately. The CPRS method, for this example and general PTA data analyses, was put forward and derived by R. van Haasteren, who also provided code to perform the sampling~\cite{decentering}. MSS achieves more efficient sampling than PRS and is consistent with CPRS. Note that the funneling issue becomes more severe when the data from multiple pulsars are used to search for a power law stochastic gravitational model, which is correlated between the pulsars. Both MSS and CPRS remain effective when applied to these more challenging sampling tasks.

\section{Discussion and Future Work}\label{sec:Future work}
The examples above illustrate the effectiveness of multi-stage sampling. MSS is able to recover the marginal distributions of hyper-parameters consistent with results from common reparameterizations, while proving more effective than naive samplers. However, the generalized hyper-model must be chosen carefully to weaken any funnels in the density so that sampling can be conducted with usual methods.

MSS is convenient when standard parameter transformations are difficult to calculate or computationally expensive to implement. For example, reparameterizations may be inefficient if a unique transformation must be computed at every step of the sampling, or are computationally expensive. Moreover, MSS is natural in systems where a generalized hyper-model already exists. The free spectral model in Section~\ref{sec:RN example} is a common spectral model in wide use today. It is the most general parameterization of a spectral model which assumes different frequency bins are statistically uncorrelated. Other diagonal spectral models (e.g. a power law model) can be inferred using MSS after an initial free spectral run.

The examples in this paper consist only of two level hierarchical models. Future work can generalize this procedure to hierarchical models with more than two levels, repeating the three steps of MSS outlined in Section~\ref{sec:intro} at every level of the hierarchy. We may also investigate the efficacy of other density estimators, such as kernel density estimation, and possible approximations. For example, assuming the generalized hyper-parameter samples are dimensionally uncorrelated improves the efficiency, but lowers the accuracy of the density estimation, which may bias the final result of MSS.

\section*{Acknowledgments}
This project was supported by National Science Foundation (NSF) Physics Frontiers Center award no. 2020265. The code used to simulate data, run analyses, and produce figures presented in this paper is publicly available at~\cite{code}. We thank Rutger van Haasteren for suggesting, deriving, and providing example code for the conditional posterior reparameterized sampling method.

\bibliography{refs}

\end{document}